\documentclass[twocolumn,floatfix,showpacs,preprintnumbers,amsmath,amssymb,prb]{revtex4}
\usepackage{graphicx,tabularx}
\usepackage{amssymb}
\usepackage[mathcal]{euscript}
\newcommand{\abs}[1]{\left\vert#1\right\vert}

\newcommand{\mgni}{$\textrm{Mg}_2\textrm{Ni}$}
\newcommand{\mgh}{$\textrm{MgH}_2$}
\newcommand{\mgcu}{$\textrm{MgCu}_2$}
\newcommand{\mgnih}{$\textrm{Mg}_2\textrm{NiH}_4$}

\newcommand{\mgfeh}{$\textrm{Mg}_2\textrm{FeH}_6$}
\newcommand{\mgcoh}{$\textrm{Mg}_2\textrm{CoH}_5$}
\newcommand{\Am}{\AA $\;$}
\newcommand{\limq}{\lim_{|\mathbf{q}|\to0}}
\newcommand{\bracet}[2]{\left<#1\vert#2\right>}

\newcommand{\bk}{\mathbf{k}}
\newcommand{\bq}{\mathbf{q}}

\begin{document}
\title{\em Ab initio \em study on the effects of transition metal doping of Mg$_2$NiH$_4$}
\author{Michiel J. van Setten}
\author{Gilles A. de Wijs}
\affiliation{Electronic Structure of Materials, Institute for Molecules and Materials, Faculty
of Science, Radboud University Nijmegen, Toernooiveld 1, 6525 ED Nijmegen, The Netherlands}
\author{Geert Brocks}
\affiliation{Computational Materials Science, Faculty of Science and Technology and MESA+ Institute
for Nanotechnology, University of Twente, P.O. Box 217, 7500 AE Enschede, The Netherlands}

\date{\today}


\begin{abstract}
\mgnih\ is a promising hydrogen storage material with fast (de)hydrogenation kinetics. Its hydrogen
desorption enthalpy, however, is too large for practical applications. In this paper we study the
effects of transition metal doping by first-principles density functional theory calculations. We
show that the hydrogen desorption enthalpy can be reduced by $\sim 0.1$~eV/H$_2$ if one
in eight Ni atoms is replaced by Cu or Fe. 
Replacing Ni by Co atoms, however, increases the hydrogen desorption enthalpy. We study the
thermodynamic stability of the dopants in the hydrogenated and dehydrogenated phases. Doping with Co or Cu
leads to marginally stable compounds, whereas doping with Fe leads to an unstable compound. The
optical response of \mgnih\ is also substantially affected by doping. The optical gap in \mgnih\ is
$\sim 1.7$ eV. Doping with Co, Fe or Cu leads to impurity bands that reduce the optical gap by up
to 0.5~eV.
\end{abstract}
\maketitle

\section{Introduction}

The large scale application of hydrogen as a fuel depends on the development of materials that can
store hydrogen in a dense form.\cite{zuttel04} Magnesium based hydrides are interesting candidate
materials for hydrogen storage, because magnesium has a low weight. The simplest hydride, MgH$_2$,
has a hydrogen storage capacity of 7.67~wt.\%. It has, however, a high hydrogen desorption
enthalpy, 0.77~eV per H$_2$, and hence an equilibrium plateau pressure ($10^{-7}$~bar) at room
temperature that is too low for practical applications.\cite{stampfer60} Moreover, the hydrogen
desorption/absorption reactions of MgH$_2$/Mg suffer from kinetic barriers, which requires
operating temperatures in excess of 570~K.
Various magnesium alloys have been studied to improve the thermodynamics and kinetics of the
hydrogen desorption/absorption
reaction.\cite{bogdanovic87,bortz98,liang99,bouaricha00,liang00,leng04}
\mgnih, which can store 3.6~wt.\% of hydrogen, has been suggested as a candidate material,
primarily because of its reasonably fast hydrogen desorpion/absorption
kinetics.\cite{reilly68,yvon81,noreus81,soubeyroux84,zolliker86-2,post87,zeng99} Several
theoretical studies have been dedicated to \mgnih.\cite{liao97,garcia99,garcia02,haussermann02,
myers02} However, its measured hydrogen desorption enthalpy of 0.70~eV per H$_2$,\cite{post87} is
barely lower than that of MgH$_2$. It leads to an equilibrium hydrogen pressure of 1~bar at
510~K.\cite{zeng99} This temperature is far too high for applications using PEM~fuel cells, for
instance.
Apart from its possible role in hydrogen storage, Mg$_2$Ni is also interesting because it can act
switchable mirror.\cite{richardson01,lohstroh04,lohstroh04-2,isidorsson02} \mgni\ is a metal,
whereas \mgnih\ is a semiconductor with a band gap of 1.7-2.0~eV.\cite{lupu93,selvam88,lupu87}
This leads to a remarkable change in the optical properties of the material upon hydrogenation and
dehydrogenation. Especially if \mgni\ is applied in thin films, the optical switching can be fast,
reversible and robust.\cite{lohstroh07} The high optical contrast opens up possibilities for \mgni\ as a hydrogen
sensing material.\cite{di05,westerwaal06,pasturel06}

For both applications the reaction enthalpy of the hydrogen desorption/absorption at typical
operating conditions is too high. Preferable would be an equilibrium hydrogen pressure of 1~bar at
room temperature. To reach this condition a hydrogen desorption enthalpy of 0.40~eV per H$_2$ is
required. A substantial amount of experimental work has been dedicated to study the effects of
doping of \mgnih\ in order to reduce its hydrogen desorption
enthalpy.\cite{darnaudery83,lei94,sun95,ikeda98,liang98,bobet00,bobet02,sato03,li04,simicic06,lohstroh07}
In this context ``doping'' means substituting a fairly large amount of Ni (or Mg) by other metals.

In this paper we report a study on the effects of doping \mgnih\ with transition metals by first
principles density functional theory (DFT) calculations. We restrict ourselves to the LT phase of
\mgnih, since the HT phase is stable only at elevated temperatures (i.e. $T>500$K) and therefore
less relevant for applications.\cite{post87} We consider substitution of nickel by cobalt, iron,
or copper in a concentration of 12.5\%, which means substituting one in eight nickel atoms. This
concentration is close to that used in recent experiments.\cite{lohstroh-pc} Our first aim is to
monitor the change in hydrogen desorption enthalpy and, in particular, to establish which dopants
(if any) lead to a reduction of the enthalpy. We show that zero point energies (ZPEs) associated
with the hydrogen phonon modes considerably influence the enthalpies. Our second objective is to
study the change in optical properties that results from doping. In particular we show that
dopants in this concentration markedly alter the dielectric function.


\section{Computational methods}

First principles DFT calculations are carried out using a plane wave basis set and the projector
augmented wave (PAW) method,\cite{paw,blo} as incorporated in the Vienna \em Ab initio \em
Simulation Package (VASP),\cite{vasp1,vasp2,vasp3}. We use the PW91 generalized gradient
approximation (GGA) for the exchange correlation functional.\cite{gga} The cell parameters are kept
at the experimental values and the atomic positions are relaxed using a conjugate gradient
algorithm. Non-linear core corrections are applied.\cite{core}

To calculate accurate reaction enthalpies for reactions involving light elements such as hydrogen,
it is essential to take into account the ZPE contribution. To calculate ZPEs we need the phonon
frequencies of the materials involved. Phonon frequencies are calculated using a direct
method,\cite{kresse95} i.e.\ the dynamical matrix is constructed from the force constants that are
obtained from finite differences. Two opposite displacements of 0.05~\Am are used for each atomic
degree of freedom. In general one needs to carry out such calculations on a super cell containing
several primitive unit cells, as the force constants do not go to zero within a distance
corresponding to a single unit cell. However, the unit cells of the materials studied in this
paper turn out to be sufficiently large, except for bulk magnesium, for which a $2\times2\times2$
super cell is used. An advantage of large unit cells is that the phonon dispersion is small. It is
therefore sufficient to calculate ZPEs from the phonon frequencies obtained at $\Gamma$.

The dielectric functions are calculated in the or independent particle random phase
approximation taking into account direct transitions from occupied to unoccupied Kohn-Sham
orbitals only. We neglect excitonic, local field and quasi-particle effects. The imaginary part of
the macroscopic dielectric function then has the form
\begin{eqnarray}\label{epsimag}
\varepsilon^{(2)}(\mathbf{\hat{q}},\omega)&=&\frac{8\pi^2 e^2}{V} \limq
\frac{1}{\abs{\bq}^2}\sum_{\bk,v,c}\\  \nonumber && \times \abs{\bracet{u_{c,\bk +
\bq}}{u_{v,\bk}}}^2 \delta(\epsilon_{c,\bk + \bq}-\epsilon_{v,\bk}-\hbar\omega),
\end{eqnarray}
where $\mathbf{\hat{q}}$ gives the direction of $\bq$; $v,\mathbf{k}$ and $c,\mathbf{k}$ label
single particle states that are occupied, unoccupied in the ground state, respectively;
$\epsilon$, $u$ are the single particle energies and the translationally invariant parts of the
wave functions, respectively; $V$ is the volume of the unit cell. Further details can be found in
Ref.~\onlinecite{kresseps}.

Almost all experimental optical data on hydrides are obtained from micro- or nano-crystalline
samples whose crystallites have a significant spread in orientation. The most relevant quantity
then is the directionally averaged dielectric function, i.e., $\varepsilon^{(2)}(\omega)$ averaged
over $\mathbf{\hat{q}}$. In this paper we only report directionally averaged dielectric functions.

The Brillouin zone integrations are performed using a modified tetrahedron
method.\cite{blochl94}
All calculations on the hydrides use a $7 \times 7 \times 7$ Monkhorst-Pack $\bf k \rm$-point mesh
for sampling the Brillouin zone, and the calculations on the metals use a $7 \times 7 \times 3$
Monkhorst-Pack $\bf k \rm$-point mesh.\cite{monk} We use $480$ bands to calculate the dielectric
function. This number of bands includes all transitions up to $30$~eV. For the materials containing
copper a plane wave kinetic energy cutoff of $341$~eV is used, and for the other materials a cutoff
of $337$~eV. To obtain accurate formation and reaction enthalpies, the total energies of all final
structures are calculated using a plane wave kinetic energy cutoff of $700$~eV.


\section{structure and stability of undoped materials: \mgnih, \mgcoh, \mgfeh, \mgni\ and elemental metals}\label{structure}

In order to assess the stability of doped \mgnih we first need the total energies of the undoped
hydrides and of all elemental metals involved. The optimized structures of \mgnih, \mgcoh, \mgfeh\
are given in Table~\ref{struc}. They are in good agreement with the experimental
structures.\cite{zolliker86,zolliker85,didisheim84} The metal atoms in LT \mgnih\ form a distorted
CaF$_2$-type structure. Four hydrogen atoms are arranged around each nickel atom in a tetrahedron.
In \mgfeh\ and \mgcoh\ the Mg and Fe/Co atoms form an undistorted CaF$_2$-type structure. In
\mgfeh\ the hydrogen atoms form regular octaeders around the iron atoms. In \mgcoh\ the hydrogen
atoms occupy five corners of a slightly distorted octahedron around each cobalt atom.

The experimental structure of \mgni\ can be found in Ref.~\onlinecite{soubeyroux84}. The unit cell
contains 12 Mg and 6 Ni atoms, which are basically hexagonally closed packed. The optimized
structure given in Table~\ref{struc} is in good agreement with experiment. For \mgh we use a
previously calculated structure.\cite{vansetten} For the elemental metals and \mgcu\ we use the
experimental lattice parameters, i.e., $a(c)=3.21(5.21)$~\AA, $a=2.87$~\AA, $a(c)=2.51(4.07)$~\AA,
$a=3.52$~\AA\, $a=3.61$~\AA\ and $a=7.03$~\AA\ for Mg, Fe, Co, Ni, Cu and \mgcu,
respectively.\cite{hcp} The magnetic elements iron, cobalt and nickel are treated by
spin-polarized calculations. The calculated pressures are small, indicating that it is unnecessary
to explicitly optimize the lattice parameters. We explicitly tested the latter for iron, since
there the external pressure was largest, and obtained an energy gain of less than 0.01~eV.

\begin{table}[!tbp]
\caption{\label{struc}Calculated atomic positions of \mgnih\ (exp.
Ref.~\onlinecite{zolliker86}), \mgcoh\ (exp. Ref.~\onlinecite{zolliker85}), \mgfeh\ (exp.
Ref.~\onlinecite{didisheim84}) and \mgni\ (exp. Ref.~\onlinecite{soubeyroux84}). In the
calculations the lattice parameters were kept at the experimental values.}
\begin{ruledtabular}
\begin{tabular}{lcccccc}
 & Space group& & &  &  &  \\
Compound  & unit cell & & & x & y & z \\
\hline
\mgnih &C2/c (15)          & Mg &$8f$& 0.2646 & 0.4863 & 0.0833 \\
& $\beta$ = 113.52$^\circ$ & Mg &$4e$& 0      & 0.0252 & 0.2500 \\
& a = 14.343~\AA           & Mg &$4e$& 0      & 0.5264 & 0.2500 \\
& b = 6.4038~\AA           & Ni &$8f$& 0.1199 & 0.2294 & 0.0801 \\
& c = 6.4830~\AA           & H  &$8f$& 0.2088 & 0.3048 & 0.3041 \\
&                          & H  &$8f$& 0.1390 & 0.3192 & 0.8760 \\
&                          & H  &$8f$& 0.0096 & 0.2908 & 0.0527 \\
&                          & H  &$8f$& 0.1243 & 0.9866 & 0.0727 \\
\\
\mgcoh &P4/nmm (129)       & Mg &$2a$& 3/4 & 1/4 & 0\\
& a = 4.463~\AA            & Mg &$2b$& 3/4 & 1/4 & 1/2 \\
& c = 6.593~\AA            & Co &$2c$& 1/4 & 1/4 & 0.2567 \\
&                          & H  &$2c$& 1/4 & 1/4 & 0.4947 \\
&                          & H  &$8j$& 0.4914 & 0.4914 & 0.2268\\
\\
\mgfeh &Fm3m (225)         & Mg &$8c$& 1/4 & 1/4 & 1/4\\
& a = 6.437~\AA            & Fe &$4a$& 0 & 0 & 0\\
&                          & H  &$24e$& 0.2425 & 0 & 0\\
\\
\mgni &P6$_2$22 (180)      & Mg &$6i$& 0.1639  & 0.3278  & 0\\
& $\gamma$ = $120^\circ$   & Mg &$6f$& 1/2 & 0 & 0.1165 \\
& a = 5.205~\AA            & Ni &$3b$&  0 & 0 & 1/2 \\
& c = 13.236~\AA           & Ni &$3d$& 1/2 & 0 & 1/2\\
\end{tabular}
\end{ruledtabular}
\end{table}

In order to obtain accurate enthalpies for reactions involving materials that contain hydrogen,
one has to take ZPEs into account. All calculated total energies and ZPEs are given in
Table~\ref{toten1}. We neglect the ZPEs of the elemental metals. The ZPE for magnesium is only
0.001~eV/atom. The ZPEs iron, cobalt, nickel and copper will be even smaller, since the atomic
weight of those elements is more than twice that of Mg. We did not calculate the ZPEs for \mgfeh
and \mgcoh, since we use these compounds only to study the stability of doped \mgnih\ with respect
to phase segregation, where the ZPE corrections are rather small.

To calculate hydrogen desorption enthalpies we also need the total energy of the hydrogen
molecule. It is calculated using a cubic cell with sides of 13~\AA. We find an equilibrium
distance of 0.7486~\AA, a vibrational frequency of 4350~cm$^{-1}$ and a dissociation energy of
4.57~eV, which compare reasonably well with the experimental values of 0.7461~\AA,
4401~cm$^{-1}$ and 4.48~eV, respectively.\cite{hcp,huber} The 0.1~eV deviation in the
dissociation energy of H$_2$ is relatively large in view of the accuracy required for calculating
hydrogen desorption enthalpies. This 0.1~eV may be considered as a correction to the reaction
enthalpies discussed below.

We calculate the ZPE for the hydrogen molecule from the energy levels of a Morse potential,
\begin{equation}\label{h}
E(n)=\hbar\omega\left(n+\frac{1}{2}\right) -
\frac{1}{4D_e}\left[\hbar\omega\left(n+\frac{1}{2}\right)\right]^2,
\end{equation}
where $\omega$ is the vibration frequency and $D_e$ is the dissociation energy. The result is given
in Table~\ref{toten1}.

\begin{table}[!tbp]
\caption{\label{toten1}Calculated total energies and ZPEs of the undoped hydrides and metals per
formula unit.}
\begin{ruledtabular}
\begin{tabular}{lrclrc}
& E (eV) & ZPE (eV) & & E (eV) & ZPE (eV)\\
\hline
H$_2$  &  $-$6.803 & 0.266 & \mgh   &  $-$8.983 &       \\
Mg     &  $-$1.524 & 0.001 & \mgni  &  $-$9.133 & 0.102 \\
Fe     &  $-$8.150 &       & \mgfeh & $-$34.511 &       \\
Co     &  $-$6.841 &       & \mgcoh & $-$29.355 &       \\
Ni     &  $-$5.459 &       & \mgnih & $-$24.053 & 0.852 \\
Cu     &  $-$3.725 &       & \mgcu  &  $-$9.45  &       \\
\end{tabular}
\end{ruledtabular}
\end{table}

\section{Doped \mgnih\ and \mgni}

\subsection{Structure}

The unit cell of the LT phase of \mgnih\ contains eight formula units. To simulate doping we
replace one of the Ni atoms by a Fe, Co, or Cu atom, thus achieving a 7:1 ratio between Ni and
dopant atoms. In simple terms one can think of undoped \mgnih\ as being constructed from Mg$^{2+}$
and (NiH$_4$)$^{4-}$ ions. The latter involve 18 valence electrons and are closed shell ions. Upon
doping it is likely that in the fully hydrogenated phase the closed shell character is maintained.
This means that (NiH$_4$)$^{4-}$ is replaced by (FeH$_6$)$^{4-}$, (CoH$_5$)$^{4-}$, or
(CuH$_3$)$^{4-}$. Thus for a Fe atom we add two extra hydrogen atoms, for a Co atom one and for a
Cu atom we remove one hydrogen atom. For all doped systems we fix the unit cell to that of undoped
\mgnih\ and we optimize the atomic positions. The external pressures on the doped systems are
small, which indicates that the gain in energy when relaxing the cell volumes will not be
significant.

The geometry of the hydrogens around the Fe and Co dopant atoms resembles the geometry in \mgfeh\
and \mgcoh\ respectively. In \mgfeh\ each Fe atom is in the center of a perfect octahedron of
hydrogen atoms with a Fe--H distance of 1.56~\AA. In Fe doped \mgnih\ the octahedron is distorted.
The H--Fe--H angles range from 80 to 100$^\circ$ and the Fe--H distances range from 1.55 to
1.58~\AA\ in case a H atom is only bonded to a Fe atom. However, four of the hydrogen atoms
surrounding a Fe atom also bond to Ni atoms, in which case the Fe--H distance is enlarged to
1.64--1.76~\AA. The hydrogen tetrahedra around such Ni atoms are distorted with Ni--H distances
from 1.51 to 1.80~\AA, whereas in undoped \mgnih they are between 1.56 and 1.58~\AA.

In the case of Co doping the distortions are much smaller. In \mgcoh\ the hydrogen atoms
surrounding each Co atom form a four-sided pyramid with the Co atom just above the basal plane of
the pyramid. To describe the geometry we denoting the basal plane hydrogens by  H$_b$ and the top
hydrogen by H$_t$. The H$_b$--Co--H$_b$ angle is 89$^\circ$, the Co--H$_b$ distance is 1.52~\AA,
the H$_b$--Co--H$_t$ angle is 97.6$^\circ$ and the Co--H$_t$ distance is 1.59~\AA. The Co-Mg
distances range from 2.75 to 2.80~\AA. The hydrogens surrounding the Co atom in doped \mgnih
form a slightly distorted pyramid, with H$_b$--Co--H$_b$ angles ranging from 83.1$^\circ$ to
94.1$^\circ$ and Co--H$_b$ distances ranging from 1.53 to 1.56\AA. The H$_b$--Co--H$_t$ angle is
93.4$^\circ$ to 105.1$^\circ$ and the Co--H$_t$ distance is 1.57\AA. The Co--Mg distances vary
from 2.69 to 2.80~\AA. The Ni--H bond lengths are not affected by Co doping.

We cannot compare the geometry of the hydrogens in Cu doped \mgnih\ to Mg$_2$CuH$_3$, since the
latter compound is not stable with respect to decomposition into \mgh\ and
\mgcu.\cite{darnaudery83} The Ni--H distances in Cu doped \mgnih\ are similar to those in undoped
\mgnih. The hydrogen atoms surrounding the Cu atom are located at three corners of a tetrahedron
with the Cu atom in the center. The Cu--H distances, 1.62 to 1.64~\AA, are slightly larger
then the Ni--H distances, 1.56 to 1.59~\AA.

\subsection{Reaction enthalpies}

In order to calculate the hydrogen desorption enthalpy of doped \mgnih\ we also need the total
energy of doped \mgni. The unit cell of \mgni\ contains six formula units per cell. If we replace
one of the Ni atoms in this cell by a dopant atom, this gives a 5:1 Ni:dopant ratio, instead of
the required 7:1 ratio. We approximate the total energy of the 7:1 ratio by the average energy of
three 5:1 doped unit cells and one undoped cell. All calculated total energies and ZPEs of the
doped hydrides and metals are given in Table~\ref{calcreshydr}.

From the data in Table~\ref{calcreshydr} we calculate the desorption enthalpy per H$_2$
molecule
\begin{eqnarray}
E_{\textrm{des}} & = &  E(\textrm{H}_2) + \frac{2}{x} \; \left[ E(\textrm{Mg}_2\textrm{Ni}_{\frac{7}{8}}\textrm{TM}_{\frac{1}{8}}) \right. \nonumber \\
&&\left. - E(\textrm{Mg}_2\textrm{Ni}_{\frac{7}{8}}\textrm{TM}_{\frac{1}{8}}\textrm{H}_x)\right],
\end{eqnarray}
where $E$(M) is the total energy of compound M and $x$ is the number of hydrogen atoms in the
hydride. The latter depends upon the dopant atom, as discussed in the previous section. The values
of $x$ are given in Table~\ref{calcreshydr}.

\begin{table}[!tbp]
\caption{\label{calcreshydr}Calculated total energies, ZPEs and hydrogen content of the doped
hydrides, Mg$_2$Ni$_{7/8}$TM$_{1/8}$H$_x$, and the metals, Mg$_2$Ni$_{7/8}$TM$_{1/8}$. The values
are per formula unit.}
\begin{ruledtabular}
\begin{tabular}{lrrr}
TM & E (eV) & ZPE (eV) & $x$ (\#H) \\
\hline
Fe &  $-$24.952 & 0.914 & 4.250 \\
Co &  $-$24.677 & 0.876 & 4.125 \\
Ni &  $-$24.053 & 0.852 & 4.000 \\
Cu &  $-$23.207 & 0.826 & 3.875 \\
\hline
Fe &  $-$9.278 & 0.100 \\
Co &  $-$9.233 & 0.100 \\
Ni &  $-$9.133 & 0.102 \\
Cu &  $-$8.890 & 0.102 \\
\end{tabular}
\end{ruledtabular}
\end{table}

The calculated desorption enthalpy of undoped \mgnih\ is 0.66~eV/H$_2$ without ZPE and
0.55~eV/H$_2$ with ZPE. The corresponding experimental value is 0.70~eV/H$_2$.\cite{post87} The
agreement is fair, if we correct for the overestimation of the H$_2$ dissociation energy,
mentioned in Sec.~\ref{structure}. Since the H$_2$ dissociation energy is overestimated by 0.1~eV,
it is reasonable to assume that the desorption enthalpy is overestimated by the same amount. This
gives an desorption enthalpy with ZPE of 0.65~eV per H$_2$, which is close to the experimental
value. The correction is a constant shift and in the following we give the uncorrected results
only.

\begin{figure}[!tbp]
\centering
\includegraphics[angle=270]{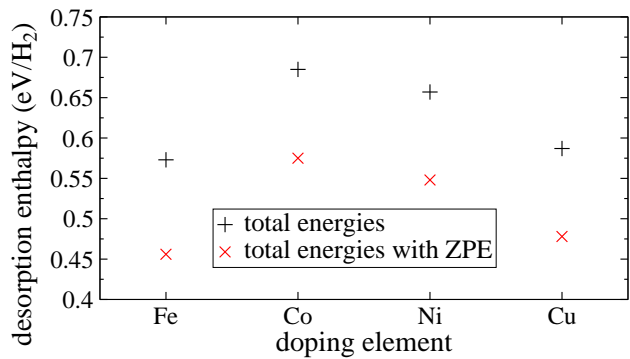}
\caption{\label{hof}(Color online) Desorption enthalpies $E_{\textrm{des}}$ (eV/H$_2$) of the
doped hydrides.}
\end{figure}

The results for $E_{\textrm{des}}$ of doped \mgnih\ are given in Fig.~\ref{hof}. These results
clearly demonstrate that the desorption enthalpy can be tuned by an appropriate doping. The
desorption enthalpy decreases considerably both for Fe and for Cu doping, i.e. by 84 and 71~meV
per H$_2$, respectively. However, Co doping increases the desorption enthalpy by 28~meV per H$_2$.
The trends for Cu and Fe doping are in qualitative agreement with experimental
data.\cite{darnaudery83,bobet02,simicic06}

Fig.~\ref{hof} also shows the calculated desorption enthalpies corrected with ZPEs. The ZPEs of
all the metals are almost identical, see Table~\ref{calcreshydr}, and the ZPEs of the hydrides
scale linearly with amount of hydrogen atoms. This means that the ZPE per hydrogen atom is almost
constant and independent of the dopant atom. Therefore, the ZPE correction to the desorption
enthalpy per H$_2$ is 0.1~eV for all compounds studied.

\subsection{Stability}

\begin{figure}[!tbp]
\centering
\includegraphics[angle=270]{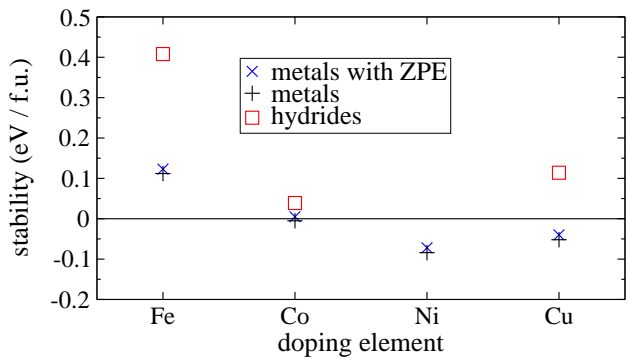}
\caption{\label{stab}(Color online) Stability with respect to phase segregation (eV / f.u.). Values
are negative if the compound is stable, see Eqs.~(\ref{reaction1})-(\ref{reaction3})}.
\end{figure}

Doped \mgnih\ is stable in thin films.\cite{lohstroh-pc} In order to assess whether kinetics plays
an important role in stabilizing these compounds, we study the thermodynamic stability of the doped
materials with respect to phase segregation. For the dehydrogenated doped \mgni\ metal we consider
decomposition into \mgni, bulk magnesium and bulk doping metal.
\begin{equation}
\textrm{Mg}_2\textrm{Ni}_{\frac{7}{8}}\textrm{TM}_{\frac{1}{8}}\rightarrow\frac{7}{8}\textrm{Mg}_2\textrm{Ni}
+ \frac{1}{8}\textrm{TM} + \frac{2}{8}\textrm{Mg}, \label{reaction1}
\end{equation}
where TM $=$ Fe, Co, or Cu. Fully hydrogenated undoped \mgnih is compared to bulk nickel and \mgh. For
the hydrogenated doped \mgnih\ we consider decomposition into phase segregated \mgnih\ and \mgfeh\ or
\mgcoh.
\begin{equation}
\textrm{Mg}_2\textrm{Ni}_{\frac{7}{8}}\textrm{TM}_{\frac{1}{8}}\textrm{H}_x\rightarrow
\frac{7}{8}\textrm{Mg}_2\textrm{NiH}_4 + \frac{1}{8}\textrm{Mg}_2\textrm{TMH}_y, \label{reaction2}
\end{equation}
with $x$ as in Table~\ref{calcreshydr},  and $y = 5,6$ for Co, Fe, respectively. Since
Mg$_2$CuH$_3$ is unstable with respect to decomposition into \mgh\ and \mgcu, we consider for the
hydrogenated Cu doped \mgnih\ the possible decomposition reaction
\begin{equation}
\textrm{Mg}_2\textrm{Ni}_{\frac{7}{8}}\textrm{Cu}_{\frac{1}{8}}\textrm{H}_x \rightarrow
\frac{7}{8}\textrm{Mg}_2\textrm{NiH}_4 + \frac{1}{16}\textrm{MgCu}_2  + \frac{3}{16}\textrm{MgH}_2.
\label{reaction3}
\end{equation}
The results are shown in Fig.~\ref{stab}.

Fe doped \mgni\ is thermodynamically unstable with respect to phase segregation into \mgni\, bulk
Mg and bulk Fe. Co doped \mgni\ is a marginally unstable material in which segregation is favored
by only $\sim 0.01$~eV per formula unit. Fe, Co doped \mgnih\ are thermodynamically unstable with
respect to segregation into \mgnih\ and \mgfeh, \mgcoh, respectively. Doping of \mgni\ with Cu
leads to a stable material. Experimental work proved the stability of Mg$_2$Ni$_{1-x}$Cu$_x$ solid
solutions;\cite{darnaudery83,li04} for $0<x<0.85$ these compounds are iso-structural with \mgni.
Experiment indicates that the hydrogenated phase decomposes into \mgh, \mgcu\ and
\mgnih.\cite{darnaudery83} This is confirmed by our calculations, see Fig.~\ref{stab}.

In conclusion, many of the doped phases are thermodynamically unstable. This does not need to
hamper their usefulness, however, since kinetics plays an important role in stabilizing the doped
compounds. The hydrogen desorption temperature lies far below the temperatures that is used to
anneal these materials.\cite{lohstroh-pc} The Cu doped \mgni\ metal is thermodynamically stable,
and hydrogenating this material can lead to a useful metastable compound.

\section{Optical properties}

\begin{figure}[!tbp]
\centering
\includegraphics[angle=270,width=8.5cm]{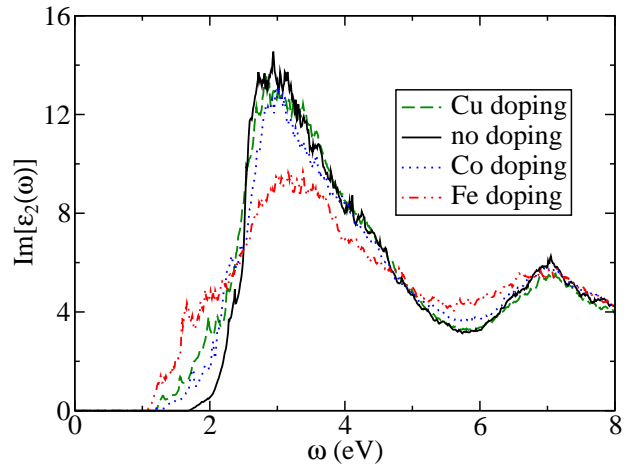}
\caption{\label{opt1}(Color online) Imaginary part of the frequency dependent dielectric functions
of doped \mgnih.}
\end{figure}

\begin{figure}[!tbp]
\centering
\includegraphics{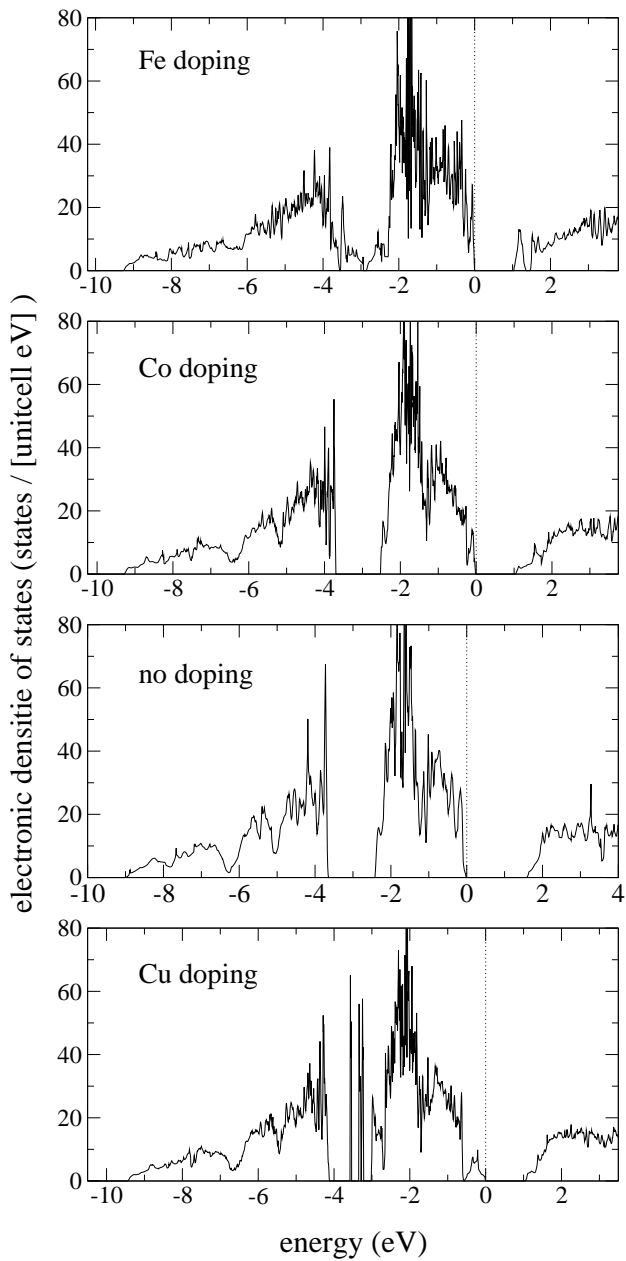}
\caption{\label{dos}Electronic densities of state of the doped and hydrides. The Fermi level is at
the top of the valence bands. The four DOSs are aligned at the bottom of the valence band.}
\end{figure}

The imaginary part of the dielectric function and the electronic density of states (DOS) of
undoped \mgnih\ are shown in Figs.~\ref{opt1} and \ref{dos}, respectively. They are in good
agreement with the results of previous calculations.\cite{myers02,haussermann02}. We find an
indirect band gap of 1.6~eV and an optical gap of 1.7~eV. This is in good agreement with the
experimental optical gap of 1.7 to 2.0~eV.\cite{lupu93,selvam88,lupu87} The agreement is in fact
remarkable since DFT usually underestimates the band gap by $30$-$50$\%. The dielectric function
of \mgnih\ has two peaks, which can be directly related to the two peaks in the DOS of the valence
bands.

The dielectric functions of doped \mgnih\ are also shown in Fig.~\ref{opt1}. Doping alters the
dielectric function and, remarkably, the size of the change correlates with the change in hydrogen
desorption enthalpy caused by the dopants, see Fig.~\ref{hof}. This can be explained by noticing
that both changes have a common cause. The changes in desorption enthalpy are due to changes in
metal--hydrogen bond lengths and bond energies. Changes in bond energies shift the energy levels
and hence can be detected in the optical spectrum. Such changes are largest in Fe doped \mgnih. An
Fe dopant atom includes two extra hydrogen atoms. It gives the largest perturbation in the \mgnih\
lattice, with almost half of the metal--hydrogen bond lengths being changed with respect to the
undoped case.

The DOS of doped \mgnih\ is given in Figure~\ref{dos}. To facilitate an internal comparison the DOS
of all compounds is aligned at the bottom of the valence band. Besides the fundamental gap between
the valence and conduction bands, we can identify a clear gap in the valence bands, between 2.4 and
3.6~eV below the Fermi level in undoped \mgnih. The states above this valence gap have a strong
metal $d$ character, whereas the lower valence states have a dominant hydrogen character. Cu and Fe
doping introduces states in the valence gap, whereas all dopants introduce states in the
fundamental gap. In the case of Cu doping these appear near the top of the valence band, whereas
for Co and Fe doping gap states appear near the bottom of the conduction band. Since we have
adjusted the amount of hydrogen upon doping, all doped materials are semiconducting.

The DOS can be used to interpret the dielectric functions. The decrease in the fundamental gap in
the DOS upon doping gives a decrease in the optical gaps. The largest changes in the dielectric
function are observed upon Fe doping. The highest peak decreases as compared to the undoped case
and the valley between the two peaks is less deep. In addition, a distinct shoulder appears at low
energy. Fe doping gives a clear peak in the DOS at the bottom of the conduction band, which yields
the distinct shoulder in the dielectric function. The two main peaks in the dielectric functions
are not shifted upon doping. This indicates that the dopants mainly give rise to additional
features via the introduction of gap states, as can be observed in the DOS. Similar conclusions
hold for the Cu and Co doped cases, but the perturbation of the \mgnih\ DOS caused by doping is
smaller than for the Fe doped case.


\section{Conclusions}

\mgnih\ is a promising hydrogen storage material with fast (de)hydrogenation kinetics. Its hydrogen
desorption enthalpy, however, is too large for practical applications. In this paper we study the
effects of transition metal doping by first-principles density functional theory calculations. We
show that the hydrogen desorption enthalpy can be reduced by 0.1~eV/H$_2$ if one
in eight Ni atoms is replaced by Cu or Fe. 
Replacing Ni by Co atoms, however, increases the hydrogen desorption enthalpy. We study the
thermodynamic stability of the dopants in the hydrogenated and dehydrogenated phases. Doping with Co or Cu
leads to marginally stable compounds, whereas doping with Fe leads to an unstable compound. The
optical response of \mgnih\ is also substantially affected by doping. The optical gap in \mgnih\ is
$\sim 1.7$ eV. Doping with Co, Fe or Cu leads to impurity bands that reduce the optical gap by up
to 0.5~eV.

We study the effects of transition metal doping on the hydrogen desorption enthalpy and the optical
properties of \mgnih\ by first-principles DFT calculations. The desorption enthalpy is reduced by
84 meV per H$_2$, if one in eight Ni atoms is replaced by an Fe atom. Replacing one in eight Ni
atoms by a Cu atom reduces the desorption enthalpy by 71 meV/H$_2$, but replacement by a Co atom
increases it by 28 meV/H$_2$. Including energy corrections due to the zero point motions of the
atoms changes the absolute values of the desorption energies by 0.1 eV/H$_2$. Since however the
zero point energies per hydrogen atom are almost independent of the compound studied, the relative
values of the desorption energies are not affected.

The thermodynamic stabilities of the doped dehydrogenated \mgni\ and the fully hydrogenated \mgnih\
compounds are studied by considering possible decomposition reactions. The results show that Cu
doped \mgni\ is stable, Co doped \mgni\ is marginally stable, and Fe doped \mgni\ is unstable with
respect to phase separation into \mgni, bulk Mg and the bulk transition metal dopant. The doped
hydrogenated \mgnih\ compounds are either marginally unstable in the case of Co or Cu doping, or, in
the case of Fe doping, clearly unstable. Kinetic barriers could be be sufficiently high to
stabilize metastable doped compounds since the hydrogen desorption temperatures are smaller than
the temperatures used to anneal these materials. Nevertheless, thermodynamics indicates that Cu is
the most promising candidate to lower the hydrogen desorption enthalpy of \mgnih.

By calculating the dielectric function within the random phase approximation we study the effects
of doping on the optical properties of \mgnih. The changes in the dielectric function can be
interpreted in terms of the electronic densities of states of the corresponding compounds. The
dopant atoms introduce states in the fundamental gap, as well as below the valence Ni $d$-band.
These states cause a shift in the onset of absorption to lower energy by up to 0.5~eV and they
decrease the relative heights of the peaks in the \mgnih\ absorption spectrum. The sizes of these
changes correlate with the change in the hydrogen desorption enthalpy caused by the dopants. Fe
doping causes the largest disruption in the \mgnih\ lattice, and the largest change in the optical
properties.

\begin{acknowledgments}
The authors wish to thank R. A. de Groot (FOM) and R. Griessen (Vrije Universiteit Amsterdam) for
helpful discussions and G. Kresse (University of Vienna) for use of the optical package. This work
is part of the research programs of `Advanced Chemical Technologies for Sustainability (ACTS)' and
the `Stichting voor Fundamenteel Onderzoek der Materie (FOM)', both financially supported by
`Nederlandse Organisatie voor Wetenschappelijk Onderzoek (NWO)'.
\end{acknowledgments}

\bibliography{dfthydrogen,mg2nih4}

\end{document}